\documentclass[twocolumn,showpacs,preprintnumbers,amsmath,amssymb]{revtex4}
\usepackage{graphicx}% Include figure files
\usepackage{dcolumn}% Align table columns on decimal point
\usepackage{bm}% bold math
\usepackage{epsfig} 

\begin{document}

\preprint{}

\title{
Comparative study of simulations of \v Cerenkov radio emission from 
high energy showers in dense media
}

\author{Jaime Alvarez-Mu\~niz}
\affiliation{Bartol Research Institute, University of Delaware,
Newark, Delaware 19716, USA
}
\author{Enrique Marqu\'es}
\affiliation{Departamento de F\'\i sica de Part\'\i culas,
Universidade de Santiago de Compostela, E-15706 Santiago 
de Compostela, Spain 
}
\author{Ricardo A. V\'azquez}
\affiliation{Departamento de F\'\i sica de Part\'\i culas,
Universidade de Santiago de Compostela, E-15706 Santiago
de Compostela, Spain
}
\author{Enrique Zas}
\affiliation{Departamento de F\'\i sica de Part\'\i culas,
Universidade de Santiago de Compostela, E-15706 Santiago
de Compostela, Spain
}

\begin{abstract} 
We compare in detail the results of simulations of electromagnetic 
showers in ice in the GeV-TeV energy range, using both the GEANT 
package and the ZHS Monte Carlo, a code specifically designed to 
calculate coherent \v Cerenkov radio pulses from electromagnetic
showers in dense media. 
The longitudinal and lateral profiles as well as the tracklengths, 
and excess tracklengths are shown to agree at the $10\%$ level. 
We briefly comment on the negligible influence of the 
Landau-Pomeran\v cuk-Migdal effect on the total shower tracklength. 
Our results are relevant for experiments exploiting
the radio \v Cerenkov technique in dense media as well as for other 
detectors that rely on the \v Cerenkov effect in water or ice.   
\end{abstract}

\pacs{96.40.Pq, 95.85.Bh, 95.85.Ry, 29.40.-n}
%PACS meaning
%96.40.Pq Extensive air showers
%95.85.Bh Radio, microwave ( >1 mm)
%95.85.Ry Neutrino, muon, pion, and other elementary particles; cosmic rays
%29.40.-n Radiation detectors

\maketitle

\section{Introduction}

Ultra high energy neutrino (UHE$\nu$) detection is one of the
experimental fields in astroparticle physics that has received most 
attention in the last two decades \cite{gaisser95}. 
Several experiments are already taking data
and several more are under way or in the proposal stage.  
Due to the low neutrino interaction
probability and the low expected neutrino fluxes, immense volumes of
detector material ($\sim 1~{\rm km}^3$ at least) are
required \cite{halzen00}. Currently explored
detection techniques exploit the observation of \v Cerenkov radiation in
the optical frequency range from neutrino-induced showers and
neutrino-induced charged leptons in dense, transparent
media \cite{halzen00,andres01,hooper02} or the search 
for horizontal air showers \cite{cronin98}. 

The observation of coherent \v Cerenkov pulses in the MHz-GHz frequency
range from neutrino induced showers in transparent, dense media, provides an
alternative method of detecting UHE$\nu$'s \cite{radhep2000}. 
The technique is most
promising at neutrino energies $>$PeV ($10^{15}$ eV) at which
effective volumes in excess of $1~{\rm km}^3$ can be achieved in a
cost-effective manner \cite{price96,frichter96}.
Electromagnetic showers are known to develop an excess 
of negative charge of about $20\%$ mainly
due to Compton scattering. 
When the wavelength of the
radiation is larger than the typical dimensions of the shower the
emission from the excess charge is coherent, 
and the power in radio waves scales as the square of shower
energy as predicted by Askary'an in the 1960's~\cite{askaryan61}. 

As interest in high energy neutrino detection grew in the mid 
1980's, proposals were made to search for radio pulses produced 
by the showers that develop when UHE neutrinos interact. 
Arrays of antennas could be used to detect the pulses 
produced in deep ice \cite{markov86} and radiotelescopes for 
those produced under the Moon surface \cite{zheleznykh89}. 
Clearly a reliable calculation of radio pulses was needed 
to explore the possibilities of these techniques. 
A fast Monte Carlo code to simulate up to PeV electromagnetic 
showers was specifically designed in the early 1990's  
to calculate the interference pattern in the MHz-GHz region  
from an electromagnetic shower in ice (the ZHS code from now on)
\cite{zas91,zas92}. 
The results 
revealed the wealth of information that is kept in the radiation 
pattern \footnote{As the radiation is coherent, it is in principle possible 
to extract the spatial charge distribution from the electric 
field amplitude.}. 
Perspectives for the technique became most encouraging for energies above 
the 10~PeV scale. 
The radio technique was further studied in the 1990's 
from three perspectives: prospects for radio 
detection arrangements were discussed \cite{frichter96}, experimental 
measurements were performed \cite{RICE01}, and the pulse simulations were 
extended to the most promising EeV range where full simulations are 
out of question with conventional computing facilities 
\cite{alvarez97,alvarez98,alvarez99,alvarez00}. 

The existence of the charge excess 
was recently confirmed in a revealing accelerator
experiment at SLAC \cite{saltzberg00}, giving a good thrust to 
the technique and generating a large number of proposals, some
of them already in operation \cite{gorham-aspen}. 
Two experiments are presently active: the RICE experiment,
an array of antennas buried in the transparent polar ice cap \cite{RICE01}
and the GLUE experiment which uses the visible side of the Moon as
target for UHE$\nu$ and cosmic ray interactions \cite{GLUE00}. New
proposals include: ANITA~\cite{gorham-aspen}, 
a balloon flown antenna looking down to the polar ice cap;   
SalSA~\cite{SalSA01}, the same concept as RICE but
exploiting the excellent optical properties of some salt domes; 
a proposal to analyze data from the FORTE satellite~\cite{FORTE} 
\footnote{FORTE searches for radio transients related 
to weather and has already triggered several million pulses in 
the 30-300 MHz frequency range}; and the LOFAR project, which will use a 
radio-astronomy array of low frequency antennas to search for radio 
emission from extensive air showers~\cite{LOFAR}. 

Radio patterns have been studied for long by an independent 
group, both from the theoretical \cite{buniy02} 
and the phenomenological sides \cite{frichter96,razzaque01}. 
Recently the predictions of ZHS have been challenged by shower 
simulations performed with the all purpose GEANT 3.21 
package which quote as much as 40\% discrepancies with ZHS in 
relevant parameters for radio-emission \cite{razzaque01}. 
The main 
discrepancy is on the absolute normalization of the amplitude of 
the electric field, a quantity that is known to be directly related 
to the difference in tracklength between electrons and positrons in 
the shower \cite{zas91,zas92}. Since the power in the signal scales with 
the square of the electric field amplitude, the discrepancy is 
unacceptable. The simulation of high energy showers in dense media is a 
cornerstone to interpret future measurements and a crucial 
tool to optimize the experiment's performance. 
Clearly the discrepancies are numerically very important, they need to be 
understood and the relative merits of alternative calculations have to 
be clearly addressed. 

In this paper we compare the ZHS code in the energy range between 100 
GeV and 10 TeV to simulations performed with both the GEANT 3.21 
package and a more recent version of GEANT, namely GEANT4. We find 
GEANT4 predictions to agree with those of ZHS to about 10\%. 
Moreover, provided that particular care is taken in defining some 
internal variables in the GEANT 3.21 package, we obtain results 
consistent with both GEANT 4 and ZHS. 
We understand that our result solves the long 
standing discrepancies between the two groups, and gives   
confidence on the calculations, which is essential for the future 
interpretation of data. Our results may be also of interest to  
neutrino experiments which are sensitive to 
tracklengths in water, 
such as AMANDA \cite{AMANDA}, BAIKAL \cite{Baikal}, ANTARES \cite{Antares}, 
SNO \cite{SNO}, SuperKamiokande \cite{SK}, 
and even to cosmic ray experiments using water \v Cerenkov tanks 
such as Auger \cite{auger}. 

This paper is organized as follows.
Firstly we discuss the basis of radio 
interference calculations and remind the reader of the 
physics involved in the coherent \v Cerenkov radio emission from showers.
In section \ref{sec:MC} we briefly discuss the inputs of 
the ZHS and GEANT simulators, and we quantitatively discuss the 
differences between the simulations. We also explore the main 
differences between the codes, namely the implementation of the LPM
effect in the GeV-TeV range and its influence on shower development.  
Section \ref{sec:conclusions} concludes the paper. 

\section{The physics of coherent \v Cerenkov radio emission and its simulation}
\label{sec:RP}

Radio experiments are sensitive to the very low 
energy electrons in the shower ($\sim 100$~keV kinetic energy) 
which contribute most to the excess charge and are responsible 
for the bulk of the coherent radio emission. 
Hybrid \cite{alvarez97,alvarez98} and analytical methods \cite{buniy02} 
have proved very useful for calculating the properties of the 
radio emission, but the complexity of the interference phenomena 
suggests a Monte Carlo approach in first instance. 

A realistic simulation of showers for radio applications
must follow all the particles explicitly down to the 100~keV-MeV kinetic
energy range. The simulations must keep 
track of the space-time positions
of the particles in the shower  
with great accuracy to be able to establish the patterns of radio emission
at MHz-GHz frequencies. 
This is a very challenging problem 
that was first approached by developing a 
specific package \cite{zas91,zas92}. 

Once all the shower particles are correctly accounted for by the 
simulations, the contributions to the electric field from all the 
individual tracks with the adequate relative phases must be calculated. 
It is convenient to work in the Fraunhofer limit with the Fourier 
time-transform of the electric field amplitude, $\vec E(\omega,\vec x)$, 
at a point $\vec x$ in the direction of the wavevector $\vec k$. 
It has been explicitly derived \cite{allan,zas92} that the contribution 
to the electric field by a charge $e$ moving at constant velocity 
$v$ for an infinitesimal 
time interval $(\rm t_1$,$\rm t_1+\delta \rm t)$ is given by: 
\begin{equation}
R \vec E(\omega,{\vec {\rm x}})=
\frac{e \mu_{\rm r}~i \omega}{2 \pi \epsilon_0 {\rm c}^2}~
{\vec v}_{\perp} \delta t ~ 
{\rm e}^{i(\omega -{\vec k} {\vec v}){\rm t_1} } 
~{\rm e}^{ikR},
\label{deltat} 
\end{equation}
where $\epsilon_0$ is the permittivity of the vacuum and $\mu_r$ the 
relative permeability of the medium, $c$ the speed of light in vacuum,
$\omega$ the angular frequency,  
and $v_{\perp}$ the projection of the particle's velocity in the direction 
perpendicular to the direction of observation. 

Eq.~(\ref{deltat}) is the basis for all the calculations of the electric 
field amplitude from showers in dense media that have been performed so far. 
Several remarks are worth emphasizing. The electric field is proportional 
to the tracklength projected in the direction perpendicular to $\vec k$. 
Eq.~(\ref{deltat}) has two phase factors. The first phase 
can be written as $(1 -n \beta \cos \theta) \omega {\rm t_1}$ 
($n$ being the refraction index and $\beta=v/c$), and takes 
care of the relative time and position at the onset of the 
time interval. The second phase is determined by the overall distance 
between the observer and the initial point of the track ($R$). 
The \v Cerenkov condition is equivalent to the first phase vanishing 
and all the points along the track contributing coherently to the electric 
field. 

In order to extend the calculation to many shower tracks, the contribution 
to the electric
field from each of them as given in Eq.~(\ref{deltat}) has to be summed, 
taking into account the different phases which are determined by  
the relative positions of the tracks. 
To apply this algorithm, individual particle tracks in a shower 
are subdivided in sufficiently small intervals.  
These should be selected so that the particle velocity is 
sufficiently constant for the approximation in Eq.~(\ref{deltat}) to be valid. 
In practice, it has been shown that it is possible to consider the average 
velocity and the end points of complete particle tracks \cite{zas92}. This 
reduces the calculation time enormously. 
In Ref.~\cite{icrc97,alvarez99} it was shown that the approximation underestimates
the amplitude at frequencies above 1 GHz but it is quite good for lower 
frequencies. 

Monte Carlo simulations performed in this way render a  
total electric field amplitude which scales accurately 
with the total tracklength in the shower, provided the second phase 
factor vanishes. This linear behavior holds for all angles of 
observation with respect to the shower axis, as long as the 
frequency is sufficiently 
low, corresponding to wavelengths $\lambda$ greater than 
the dimensions of the shower. At the 
\v Cerenkov angle the proportionality extends to frequencies
$\sim$ GHz. 
Above the GHz range however random phases destroy the scaling behavior. 
This is mainly due to the lateral spread of the shower but there are 
also effects associated with particle trajectories not aligned with 
shower axis and to time delays. 

It is helpful to think of the angular distribution of
the \v Cerenkov pulse as the Fourier transform of the 
longitudinal development of the excess charge \cite{alvarez99}.
The main {\sl diffraction peak} appears at the
\v Cerenkov angle. As the shower becomes longer, 
the angular width of the pulse becomes smaller.
This is particularly important at energies above $\sim$PeV in ice, 
where electromagnetic showers suffer the Landau-Pomeran\v cuk-Migdal 
effect \cite{LPM} and become very elongated. As a consequence
the angular distribution of the radio pulses becomes very sharply
peaked \cite{alvarez99,alvarez00}.

As a final remark, Eq. (\ref{deltat}) 
corresponds to the Fraunhofer 
condition which applies when   
($\lambda~v \delta {\rm t}) << R^2$. Eq. (\ref{deltat}) is clearly valid 
for {\it any R} provided the time interval is chosen sufficiently short. 
It is possible to apply this expression to the Fresnel region, one only 
has to ensure that the relative phase factors between tracks 
are chosen adequately. This method was used in 
reference \cite{alvarez00} to calculate the electric field
at small distances to the shower.  

\section{Simulation of electromagnetic showers: ZHS vs GEANT} 
\label{sec:MC}

A reliable Monte Carlo (MC) code that simulates shower particles in detail 
is an essential intermediate step to estimate the expected radio-emission 
from a high energy shower. Several options can be used 
for this purpose \cite{zas92,EGS4,FLUKA,GEANT,GEANT4} and here we 
concentrate on two of them namely, 
GEANT in the two versions: GEANT 3.21 \cite{GEANT} and GEANT4 \cite{GEANT4} 
and the ZHS code \cite{zas92}. We briefly discuss 
the most relevant processes considered by both. Further details can 
be obtained from the references.

The ZHS is a fast code optimized for the simulation 
of electromagnetic showers in homogeneous ice. It 
follows electrons, positrons, and photons until an externally fixed 
energy threshold is reached. 
The program accounts for bremsstrahlung and pair production, taking 
screening corrections into consideration as well as the 
Landau-Pomeran\v cuck-Migdal (LPM) effect \cite{LPM}. 
It also accounts for multiple elastic scattering, M\"oller, Bhabha, 
and Compton scattering, as well as electron-positron annihilation. 
The time delays of the particles with respect to an ideal particle moving
along the shower axis at the speed of light are carefully implemented, 
accounting for subluminal velocities, geometrical effects, and 
multiple elastic scattering. The longitudinal profile of 
electromagnetic showers obtained with ZHS was shown to agree with standard 
parameterizations of the depth development in \cite{zas92}.
In \cite{radhep2000,icrc01} ZHS was also shown to agree with
standard parameterizations of the lateral distribution of 
electrons and positrons in electromagnetic showers. 

GEANT is a well known, well tested and widely used simulation
and detection package which is suitable for a large range of 
applications. Version 4 is a recent update with several 
improvements. 
Many computational parameters such as energy
thresholds, or the minimum step length in
particle propagation, and even cross sections
and corrections may be included or not by the user.
However, care has to be taken in setting input parameters which
are consistent and in selecting the appropiate physical processes to the
problem in consideration \cite{GEANT}.

\subsection{ZHS vs GEANT3: Discrepancies}

The most important quantities in the calculation of the coherent radiopulses
are the total, total projected, and excess tracklengths \cite{zas92}.
The total tracklength is defined as the sum of the lengths of all charged 
particle trajectories in the shower.
The total projected tracklength is the sum of the projections of the particle 
paths along the shower axis. The excess charge track length is 
simply the difference between positive and negative charged 
projected tracklengths. They are calculated
in the ZHS code by default. These magnitudes 
have been found to be linearly dependent on the primary energy to a great
degree of accuracy, and are directly proportional to the peak value
of the electric field at the \v Cerenkov angle. This is just a consequence 
of energy conservation in the shower. In fact it has been shown that 
shower fluctuations do not affect the total tracklength and that by 
looking at the peak value of the coherent electric field, an accurate 
measurement of the shower energy can be obtained 
\cite{zas92,alvarez97,alvarez98}\footnote{Certainly fluctuations in tracklength 
are much smaller than uncertainties associated to cross sections and  
corrections according to different authors.}.
 
Simulations carried out with GEANT 3.21 in \cite{razzaque01} predict a
total, total projected and excess projected tracklengths which are
between $20\%$ and $40\%$ smaller than the corresponding values obtained
with ZHS.
The number of particles at shower maximum in \cite{razzaque01} is about 
$35\%$ smaller than for the ZHS. In the following we address the origin of the
discrepancies and establish that the simulations in \cite{razzaque01}
underestimate the value of these crucial parameters. 
 
We have simulated showers of different energies using ZHS, GEANT 3.21
and GEANT 4 and studied their depth development, lateral distribution
and the three tracklengths which are relevant for radio calculations. 
In order to
understand the results of \cite{razzaque01} we have made two different
sets of simulations changing an internal parameter of GEANT 3.21.
The parameter is called IABAN and it is discussed in more detail 
in the appendix. Here it should be sufficient
to say that changing the value of IABAN is equivalent to 
fixing different values of the external energy threshold below which
electrons are no longer followed in the simulations.
This, we believe, is the ultimate cause of the discrepancies 
in the tracklength between ZHS and the results presented in \cite{razzaque01}. 
We perform two sets of
simulations with GEANT 3.21: one corresponding to the default values, set (a), 
and the other using a modified setting which correctly accounts for all
the electrons in the shower which are above the kinetic energy
threshold, set (b), (see the appendix for details).
 
\begin{table*}
\caption{Total, total projected and excess projected tracklength in ice
as obtained in GEANT 3.21, GEANT 4 and ZHS. The results show the
average over 100 electron-induced showers of primary energy 100 GeV.
The electron kinetic energy threshold is 100 keV (0.611 MeV
total energy) in all simulations. GEANT 3.21 simulations were performed
with the following values of IABAN (see appendix): (a) IABAN=1 and (b) IABAN=2.
The numbers in parenthesis indicate the relative differences (in $\%$)
taking the GEANT4 results as reference.
\label{table:track}
}
\renewcommand{\arraystretch}{1.5}
\begin{tabular}{ c | c | c | c | c }
\hline
~MC Code~&~GEANT3~(a)~&~GEANT3~(b)~&~GEANT4~&~ZHS~\\ \hline
~Total track~[m]~&~ 409.2~(-30)~&~ 577.9~(-2)~&~ 587.9~(0)~&~ 642.3~(9)~\\
~Total projected~[m]~&~ 372.9~(-18)~&~ 450.0~(-1)~&~ 453.2~(0)~&~ 516.7~(14)~\\
~Excess projected~[m]~&~ 92.4~(-25)~&~ 123.5~(-1)~&~ 122.7~(0)~&~ 132.4~(8)~\\
~Excess/Total~& 0.225 (+8) & 0.214 (+2) & 0.208 (0) & 0.206 (-1) \\
\hline
\end{tabular}
\end{table*}

Table~\ref{table:track} summarizes the numerical values of the
total tracklength, total projected tracklength, and excess
projected tracklength in ice as obtained in GEANT 3.21, GEANT 4,
and ZHS. We averaged the tracklength over 100 electron-induced showers
of energy 100 GeV. The electron kinetic energy
threshold is $K_{\rm th}$=100 keV (or $E_{\rm th}=0.611$ MeV
total energy) in all simulations.
Using the default version of GEANT, GEANT 3.21 (a), we reproduce the
results of~\cite{razzaque01} to the statistical accuracy of the 
simulations. With the setting GEANT 3.21 (b), which accounts 
for an external threshold $E_{\rm th}=0.611$~MeV
consistent with what is explicitly done in the ZHS MC, 
the $40\%$ discrepancies between ~\cite{razzaque01} and ZHS 
on the tracklength as well as on the number of particles at 
shower maximum, are reduced to $10\%$. Furthermore the results of ZHS,
GEANT3 (b) and GEANT 4 are within $10\%$ of each other.
In Ref.~\cite{razzaque01} it is argued that an upper bound to the tracklength 
per unit energy is simply obtained by calculating 
the inverse of the minimum energy loss of an electron in ice 
$[dE/dX]_{\rm min}^{-1}$. We remark here that this 
is indeed an upper bound to the average length traveled by a single electron, 
however it does not take into consideration the contributions to shower 
tracklength through secondary electrons produced by bremsstrahlung photons 
that contribute to the energy loss of the primary electron. 

\begin{figure}[ht]
\centerline{
\mbox{\epsfig{figure=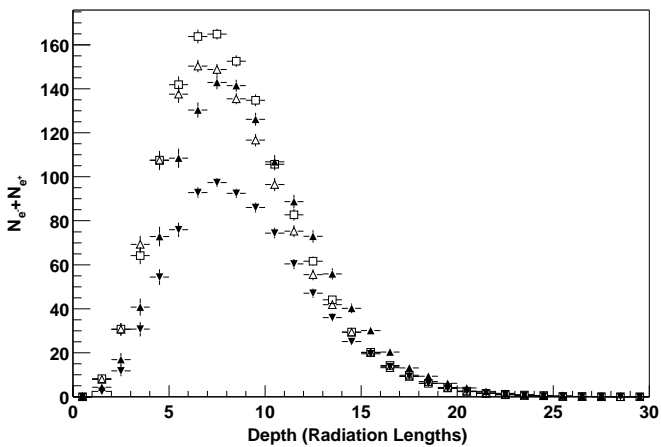,width=6.5cm,angle=-90}}
}
\caption{Average number of electrons plus positrons in 100 GeV electron
initiated showers, averaged over 100 showers, as a function of depth in ice.
The profiles were obtained with different Monte Carlo programs; 
ZHS (squares), GEANT 4
(unfilled triangles), GEANT 3 with IABAN=2 (filled triangles),
GEANT 3 with IABAN=1 (filled inverse triangles).
Depth is measured in radiation lengths ( 1 r.l.= 36.08 g/cm$^2$).}
\label{fig:shlong}
\end{figure}
 
\begin{figure}[ht]
\centerline{
\mbox{\epsfig{figure=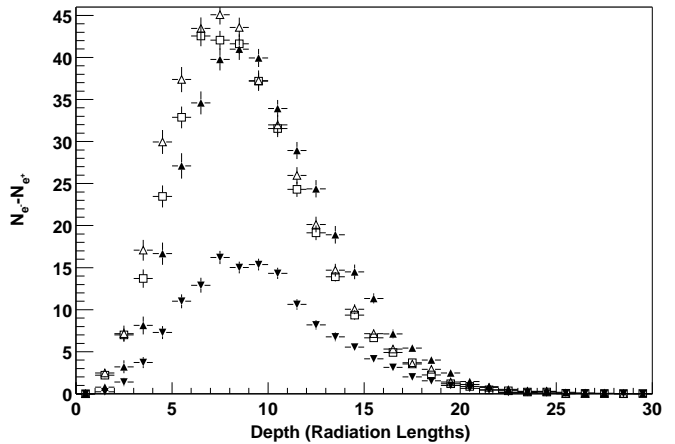,width=6.5cm,angle=-90}}
}
\caption{Same as Fig.~\ref{fig:shlong} for the average number of electrons 
minus positrons.} 
\label{fig:shlong-excess}
\end{figure}

\begin{figure}[ht]
\centerline{
\mbox{\epsfig{figure=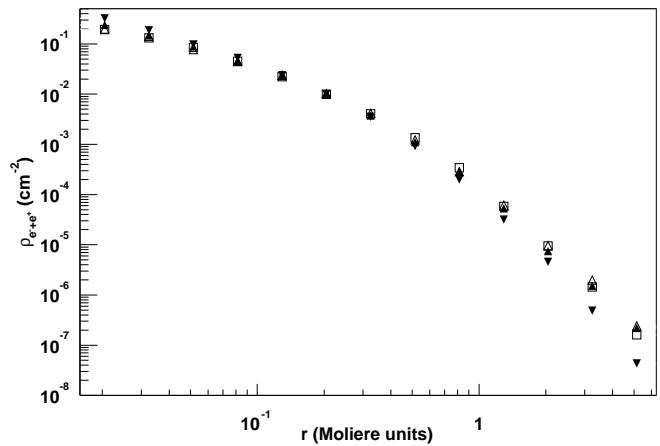,width=6.5cm,angle=-90}}
}
\caption{Lateral distribution of electrons and positrons in 100 GeV showers
in ice at a depth of 7 radiation lengths as obtained with different MC codes.
ZHS (squares), GEANT 4
(unfilled triangles), GEANT 3 with IABAN=2 (filled triangles),
GEANT 3 with IABAN=1 (filled inverse triangles). All densities are normalized
to the same number of particles. 1 Moli\`ere unit is 10.4 g/cm$^2$ in ice.}
\label{fig:ld7X0}
\end{figure}

\begin{figure}[ht]
\centerline{
\mbox{\epsfig{figure=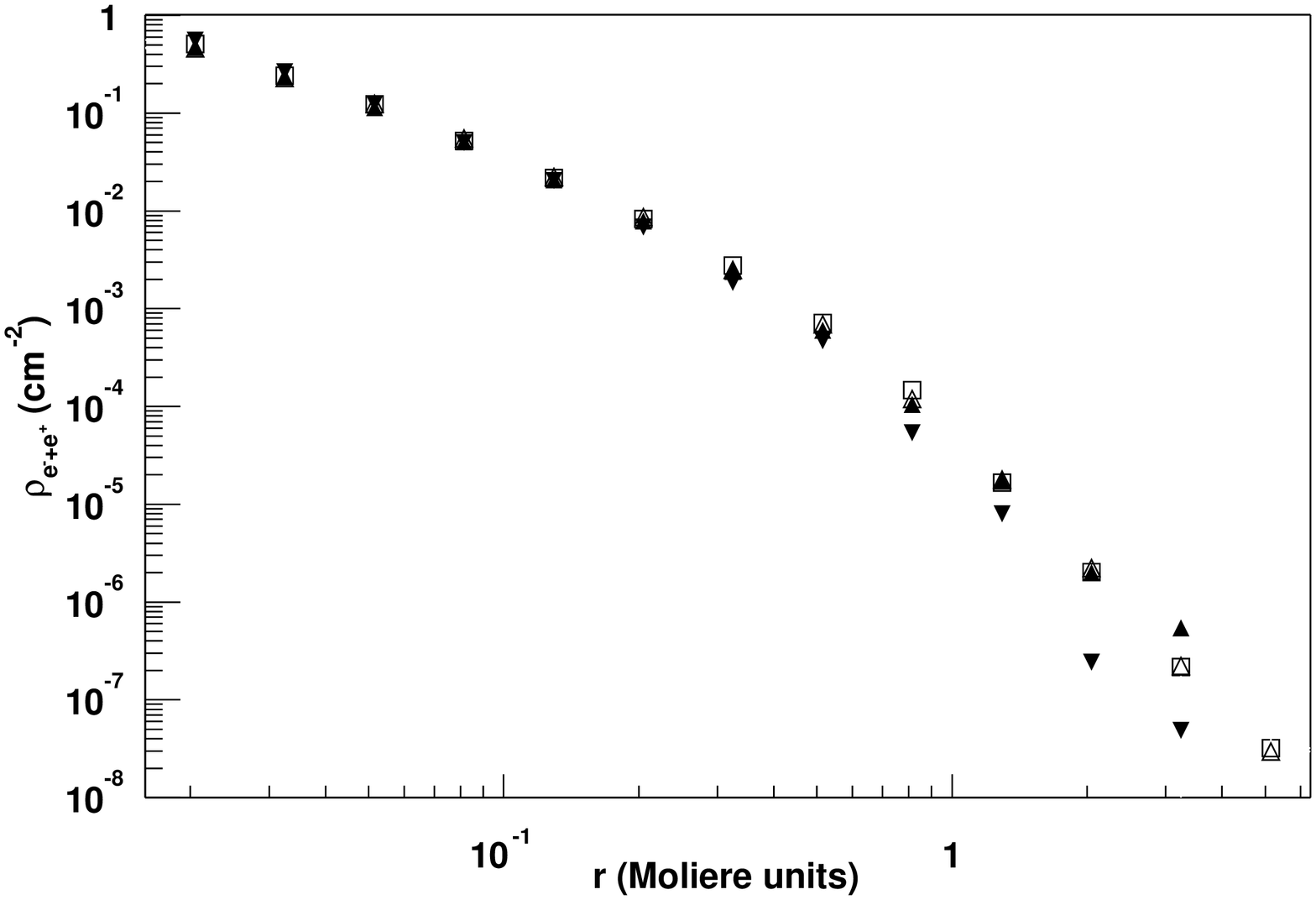,width=6.5cm,angle=-90}}
}
\caption{Same as figure \ref{fig:ld7X0} at a depth of 4 radiation lengths.}
\label{fig:ld4X0}
\end{figure}

\begin{figure}[ht]
\centerline{
\mbox{\epsfig{figure=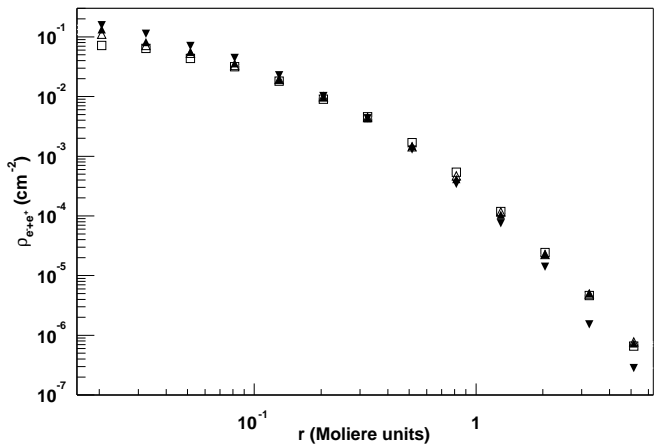,width=6.5cm,angle=-90}}
}
\caption{Same as figure \ref{fig:ld7X0} at a depth of 10 radiation lengths.}
\label{fig:ld10X0}
\end{figure}

Fig.~\ref{fig:shlong} (Fig.~\ref{fig:shlong-excess}) shows 
the average longitudinal development of the total (excess) charge  
for 100 electron-initiated showers of energy 100 GeV in ice. The
same MC codes as in table~\ref{table:track} were used to produce
the plots. The number of particles at shower maximum as obtained with 
GEANT 3.21 (a) (see table caption and appendix) agrees with the result quoted  
in~\cite{razzaque01}, and
differs by $30\%$ from the value predicted by ZHS. However the expectations
from GEANT 3.21 (b), GEANT4, and ZHS are again within $\sim 10\%$ of each
other. In Figs.~\ref{fig:ld7X0},~\ref{fig:ld4X0} and \ref{fig:ld10X0}
we show the lateral distribution of electrons and positrons at shower
maximum corresponding to $\sim 7 $ radiation lengths 
($1~{\rm r.l.} = 36.08~{\rm g/cm^2}$), as well as  
at 4 r.l., and at 10 r.l. for the four nominal MC codes.
Again, GEANT 4, ZHS, and GEANT 3.21 (b) give similar results
whereas GEANT 3.21 (a) underestimates the total number of particles
and also slightly distorts the lateral distribution. 
 
The discrepancies between the radio emission simulations presented in
our previous work \cite{zas91,zas92,alvarez97,alvarez98,alvarez99,alvarez00} 
and those reported in \cite{razzaque01} can be mostly
interpreted in terms of the normalization of the tracklength.  
The remaining differences can be attributed
to the effect of track subdivision. If small subintervals are chosen, the
overall behavior obtained by ZHS up to 5~GHz is very similar to that
shown in ref.\cite{razzaque01}, provided the normalization is corrected
according to the tracklength deficit. It is however difficult to be confident
on the results at higher frequencies, because very small 
relative time differences of only 0.02~ns and distances between particles 
of the order of 6 mm, are expected to have an important effect on  
the interference patterns. Theoretical calculations based
on simple current density parameterizations which neglect time delays and
front curvature \cite{alvarez00,buniy02} are not expected to be useful
in this frequency range for the same reason.

The good agreement between GEANT 3.21 (b), GEANT 4 and ZHS 
is a remarkable achievement given the fact that
ZHS and GEANT have been developed independently of each
other and are of very different nature and purpose. 
There are two important 
differences between the codes but a detailed
analysis is not the purpose of this paper. In the next
subsection we show that none of them can account for 
a $50\%$ discrepancy in the tracklength, although they might
explain the remaining ($\sim$ 10\%) discrepancies in the observables.

Finally we would like to stress that the results presented above are
all reproducible with the available (upon request) codes \footnote{Please
send requests to {\bf alvarez@bartol.udel.edu}}. 
We believe we have solved this long standing and crucial discrepancy 
between the different simulations. The implications for radio projects
are obvious but it may be stressed that other simulations relying
on tracklength in water detectors that have been made for many other
experiments such as those computing \v Cerenkov light in water/ice, 
should be checked against the results presented in this paper.

\subsection{ZHS vs GEANT: Differences}

It should be remarked that both codes take the same interactions into account, 
except for two important differences, namely the LPM corrections which are not 
accounted for in GEANT 3.21 \cite{GEANT4}, and the photoelectric effect which 
is ignored in the ZHS code. 
The photoelectric effect is not expected 
to be important for low $Z$ nuclei such as hydrogen and oxygen 
for energies above the electron \v Cerenkov energy threshold 
($\sim 100$ keV kinetic energy in ice). We will discuss the 
LPM effect and show that 
it hardly represents any effect for the tracklength calculations.

The LPM effect stems from the fact that the 
interaction distance in bremsstrahlung or pair production
is inversely related to the momentum transferred to the nucleus ($q$).
When the initial and final electron momenta in bremsstrahlung (or 
the electron and positron momenta in pair production) become 
ultrarelativistic, the two electron momenta are almost collinear and
the longitudinal momentum transfer $q_{\vert\vert}$ 
can be very small. Conversely,
the distance along which the interaction occurs
might become large, comparable to the interatomic spacing which 
depends on the density of the medium. Under these circumstances the electron  
encounters additional atoms along the interaction distance which
cause multiple Coulomb scattering. This introduces destructive
interference in 
the interaction matrix element and as a result the 
total bremsstrahlung and pair production cross sections are 
suppressed~\cite{LPM}.
The suppression becomes important when the average multiple
Coulomb scattering angle ($\theta_s$) is comparable to the scattering 
angle in the bremsstrahlung or pair production interactions ($\theta_i$).

In pair production the two final electron momenta are collinear
only when the photon energy is   
above an energy scale ($E_{\rm LPM}$) which decreases with  
the density of the medium ($E_{\rm LPM}\sim 2$ PeV in ice). 
The same occurs in bremsstrahlung  
when the energy of the electron is above $E_{\rm LPM}$, but also when 
the fraction of the electron's energy carried away by the photon is
small.  
    
Following Ref.~\cite{stanev82} a parameter ($s$) is introduced
through the condition $\theta_s\sim \theta_i$. 
When $s<1$ the LPM has an important effect on bremsstrahlung
and pair production interactions. $s$ is given by~\cite{stanev82}:

\begin{eqnarray}
s\propto \sqrt{\frac{E_{\rm LPM}}{E}~ \frac{u}{1-u}}~~~~~~{\rm bremsstrahlung},\\
s\propto \sqrt{\frac{E_{\rm LPM}}{E}~ \frac{1}{v(1-v)}}~~~~~~{\rm pair~~production},
\end{eqnarray}
where $u$ is the fraction of the energy of the electron 
carried by the radiated photon in bremsstrahlung, 
and $v$ is the fraction of the photon's
energy carried by the electron or the positron in 
pair production. Clearly $s<1$ in both equations when $E \gg E_{\rm LPM}$  
but also when $u$ is smaller than $u_{\rm LPM}\simeq E/E_{\rm LPM}$ 
in bremsstrahlung.

As a consequence, it can be easily shown that 
the differential bremsstrahlung cross section scales 
as $dN/dk\propto 1/\sqrt k$ instead
of exhibiting the Bethe-Heitler behavior 
$dN/dk\propto 1/k$, where $k$ is the energy of
the photon. The integral of the differential cross section 
is naturally regularized, preventing the 
infrared catastrophe at very low photon energies.
Moreover, the average fraction of the electron's energy
carried away by bremsstrahlung photons $\langle u\rangle$ 
increases with energy due to the suppression  
of photons with energy below $u_{\rm LPM} E$. 
This effect has been experimentally measured at energies $\sim 10-25$ GeV
in a variety of materials~\cite{cavalli-sforza95,cavalli-sforza97}
(see~\cite{klein99} for a comprehensive review). 

\begin{figure}[ht]
\centerline{
\includegraphics[width=8.5cm]{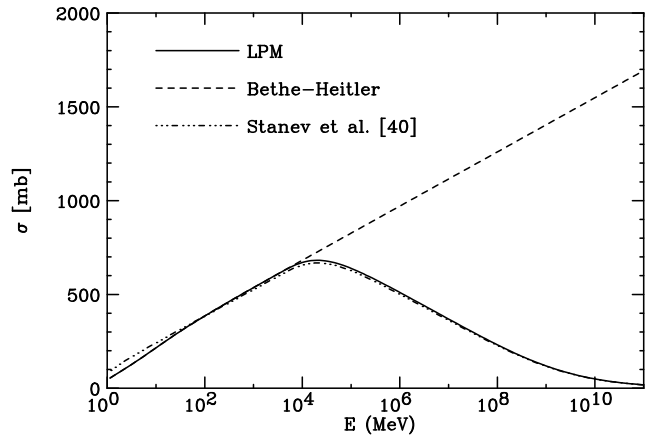}
}
\caption{Total bremsstrahlung cross section as a function
of electron energy. The solid line is the total cross section
when the LPM effect is accounted for. The dashed line is 
the Bethe-Heitler cross section i.e. without the LPM effect.
The dashed-dotted line is the total cross section as obtained
integrating the differential cross section in reference~\cite{stanev82}.
The cutoff photon energy is 100 keV for the three curves.
}
\label{fig:bremscross}
\end{figure}

Fig.~\ref{fig:bremscross} shows the bremsstrahlung cross section 
in ice when the LPM effect is considered and the 
well known Bethe-Heitler cross section (without LPM effects). 
Despite the fact that $E_{\rm LPM}$ is in the PeV energy range
the suppression of the bremsstrahlung cross section starts already at  
around 10 GeV due to the effect explained above. 
The LPM bremsstrahlung cross section drops as $\sqrt{E}$ in contrast
to the logarithmic increase with energy of the 
Bethe-Heitler cross section. Also shown for comparison is the 
LPM cross section obtained in \cite{stanev82}. The small 
differences between both LPM cross sections can be attributed to the 
Koch and Motz empirical low energy corrections \cite{koch} considered 
in ZHS but ignored in \cite{stanev82}.

Shower development is nevertheless fairly insensitive to 
the large differences between the LPM and the Bethe-Heitler
bremsstrahlung cross section in the energy range below 100 TeV.   
We have calculated the electron energy loss per radiation length in the 
100 GeV - 100 TeV energy and we have seen that it 
is unaffected by the inclusion or not of the
LPM effect. The reason for this is that the 
decrease in the bremsstrahlung cross section in this energy range,
is compensated by the expected increase in $\langle u\rangle$, so that  
the product $\langle u\rangle \sigma$ (i.e. the energy loss) 
is the same as in the Bethe-Heitler calculation.  
In contrast, at energies above 
$E_{\rm LPM}$ this cancellation does no longer exist and the 
energy loss including the LPM effect is lower than the one predicted
by the Bethe-Heitler cross section. This added to the fact that
the pair production cross section is also suppressed above $E_{\rm LPM}$,
leads to the characteristic elongated longitudinal profiles of 
LPM electromagnetic showers \cite{alvarez97,misaki}. 

Finally it is worth remarking that the total tracklength is completely
insensitive to the LPM effect at all energies. This is just a manifestation
of energy conservation in the shower as explained before.  

There are some other minor differences in the treatments 
and approximations used for the cross sections for relevant electromagnetic processes
which have a slightly different implementation in GEANT and ZHS. 
The treatment of Moli\`ere multiple elastic scattering is also different. 
However in Ref.~\cite{razzaque01} it is shown that
a small variation in these cross sections by as much as $25\%$ cannot be 
responsible for large discrepancies in shower parameters. 
Surely these minor differences together with the LPM effect are likely to be 
responsible for the remaining $10\%$ differences between ZHS, GEANT 3.21 (b) 
and GEANT 4.  

\section{Conclusions}
\label{sec:conclusions}

We have simulated electromagnetic showers in ice 
using two different Monte Carlo codes, GEANT and ZHS. 
The discrepancies discussed in reference \cite{razzaque01} between 
the ZHS code and GEANT diminish to the 10 \% level  
when the GEANT 4 version is used, 
and also when the GEANT 3.21 code is carefully set to allow for a  
constant energy threshold. 

We have shown that the influence of the LPM effect  
on shower development is negligible for energies below 
$E_{\rm LPM}$, despite the fact that the bremsstrahlung 
cross section is significantly affected. As it has been now
well established, the shower longitudinal profile 
is dramatically affected by the LPM at energies
above $E_{\rm LPM}$.  

It is worth stressing that ZHS is a Monte Carlo 
specifically designed to calculate coherent radiopulses 
in electromagnetic showers. On the other hand, GEANT is a general purpose 
Monte Carlo designed to deal with a large variety of problems. 
The fact that both unrelated and independent programs give 
results to the $10\%$ level of precision gives us confidence on 
both packages.
The ZHS program is also suitable for simulations of experiments that use 
ice or water and the \v Cerenkov imaging technique, as long as muons are not
important. It can also be extended to other media. 
In addition, ZHS is considerably faster than the GEANT code and
can reach much higher energies, making it adequate for 
high energy neutrino experiments. 

Small discrepancies that remain between ZHS and GEANT are possibly due to minor 
differences which can be attributed to physical uncertainties in the 
processes involved. 
As a result all previously published calculations based on the ZHS Monte Carlo 
are expected to be correct to the same degree of accuracy, modulo other 
uncertainties which were discussed in the original papers. 

\begin{acknowledgments}
We thank D.Z. Besson, D.W. McKay, J.P. Ralston, S. Razzaque, D. Seckel
and S. Seunarine for sharing their results with us and for 
discussions. We also thank 
R. Engel, F. Halzen and T. Stanev for many stimulating 
discussions on this topic.  
This work is supported by Xunta de Galicia (PGIDT00PXI20615PR), by 
CICYT (AEN99-0589-C02-02), and by MCYT (FPA 2001-3837). J.A.-M. is 
supported by NASA Grant NAG5-7009. R.A.V. is supported by the
``Ram\'on y Cajal'' program.
\end{acknowledgments}

\appendix
\section{The GEANT setting IABAN}
 
The default setting of GEANT includes
a variable named IABAN \footnote{Documentation about this variable
can be found in the comments of the gtelec.F subroutine inside the
GEANT code. See line 370 of the subroutine provided in the
GEANT 3.21 distribution. Unfortunately it is not documented in the
manual pages of GEANT.} whose value is set to 1. The effect of
this setting is to stop tracking a fraction of the electrons and
positrons of energy below 10 MeV,
even when the external electron energy threshold is set to a lower value.
To be more precise, when the distance to the next bremsstrahlung
interaction is larger than the range of the electron, the electron is
stopped and its tracklength is not computed. Typically in a
100 GeV shower $\sim 50\%$ of the electrons
of energy below 10 MeV are stopped and eliminated
from the simulation. The stopped electrons
account for $40\%$ of the total tracklength.
As a consequence
the total, total projected and excess projected
tracklengths and hence the
total electric field emitted by a shower are underestimated
by $\sim 40\%$.
 
If the variable IABAN is set to 2, all the electrons and positrons
in a GEANT simulation are tracked down to the external
energy threshold which is chosen by the user. This 
setting gives the correct tracklength.
 
In GEANT4 the switch IABAN does no longer exist, and particles
are always correctly tracked down to the external threshold.
As a consequence all the particles that build up the
radio-emission are correctly accounted for.

%%%%%%%%%%%%%%%%%%%%%%%%%%%%%%%%%%%%%%%%%%%%%%%%%%%%%%

%%%%%%%%%%%%%%%%%%%%%%%%%%%%%%%%%%%%%%%

\end{document}